# Persistence and Big Data Analytics Architectures for Smart Connected Vehicles

Sebastian Scholze
Institut für angewandte Systemtechnik
Bremen GmbH
Bremen, Germany
scholze@atb-bremen.de

Fulya Feryal Horozal
Institut für angewandte Systemtechnik
Bremen GmbH
Bremen, Germany
horozal@atb-bremen.de

Marie-Saphira Flug
Institut für angewandte Systemtechnik
Bremen GmbH
Bremen, Germany
flug@atb-bremen.de

Ana Teresa Correia
Institut für angewandte Systemtechnik
Bremen GmbH
Bremen, Germany
correia@atb-bremen.de

*Abstract*— Up until recently, relational databases were considered as the de-facto technology for persisting and managing large volumes of data. This came to change with the emergence of enterprises producing extremely large datasets and having unprecedentedly high availability requirements. The need for levels of availability beyond those supported by relational databases and the challenges involved in scaling such databases horizontally led to the emergence of a new generation of purpose-specific databases grouped under the term NoSQL. The NoSQL databases despite being designed with horizontal scalability as a primary concern and deliver increased availability and fault-tolerance, they have the cost of temporary inconsistency and reduced durability of data. The present paper presents a concept and application of how TYPHON aims to face the challenges emerging from the attempts to balance the best world of relational and NoSQL databases, by supporting the migration towards hybrid data persistence architectures.

## I. INTRODUCTION

As described in [1], up until recently, relational databases were considered as the de-facto technology for persisting and managing large volumes of data. This came to change with the emergence of enterprises such as Google, Twitter, Facebook, Amazon etc. which were faced with extremely large datasets and unprecedentedly high availability requirements. The need for levels of availability beyond those supported by relational databases and the challenges involved in scaling such databases horizontally led to the emergence of a new generation of purpose-specific databases grouped under the term NoSQL. In general, NoSQL databases are designed with horizontal scalability as a primary concern and deliver increased availability and fault-tolerance at a cost of temporary inconsistency and reduced durability of data [2].

To balance requirements for data consistency and availability, organisations increasingly migrate towards hybrid data persistence architectures comprising both relational and NoSQL databases for managing different subsets of their data. The consensus is that this trend will only become stronger in the future; critical data will continue to be stored in ACID (predominately relational) databases while non-critical data with high availability requirements will be progressively migrated to purpose-specific NoSQL databases. Moreover, as the volume and the value of textual content constantly grows, built-in support for sophisticated text processing in data persistence architectures is increasingly becoming essential.

The approach presented in this paper is part of a wider research. The overall concept for this research is briefly presented in Section II. In the following sections, the paper focusses on the approach, concept and prototype implementation of technologies for deploying polystores. Section III gives a short overview on different deployment technologies, while Section IV introduces the concept for a hybrid polystore deployment modelling language. In Section V the prototype implementation is described and Section VI show a real world application scenario of the presented approach. Finally, the paper is concluded in Section VII.

## II. PROPOSED CONCEPT

Fig. 1 shows an overview of the proposed architecture. The process starts with the creation of a model of the polystore. Developers, using a textual and graphical Domain Specific Language (DSL) called TyphonML, create models that include information regarding the concepts appearing in the polystore, their attributes and their relationships. These models, labelled as "TyphonML models" in Fig. 1, also include information about the databases that will be involved in the system. As a result, they represent the high-level infrastructure of a hybrid polystore [3].

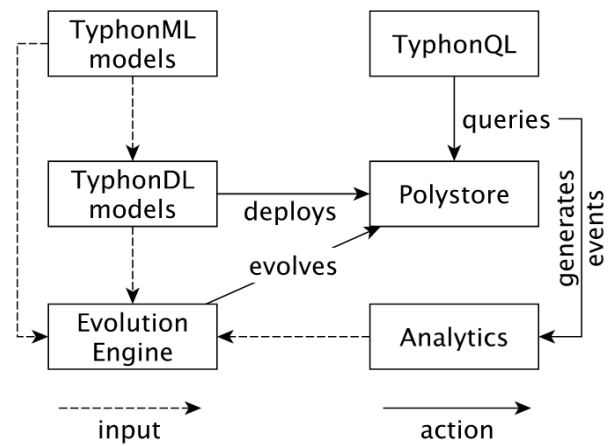

Fig. 1. Overview of proposed architecture

To bridge the gap between high-level TyphonML models and ready-to-use polystores, an intermediate polystore deployment modelling language (TyphonDL) is used. TyphonDL provides concepts that lie at an abstraction level between that of TyphonML, and that of specific data stores and virtual machine configuration technologies. TyphonML models are transformed to "TyphonDL models" and are enhanced with more fine-grained database-specific options. TyphonDL models represent the deployment infrastructure of that polystore in terms of the specific cloud platform and deployment tools employed and are used to generate the necessary installation and configuration facilities that, when executed, can assemble the Polystore in an automated manner.

As data will be distributed across a number of heterogeneous databases a common data manipulation language is used. "TyphonQL" is developed for performing data manipulation and query commands (e.g. insert, delete, etc.) [4]. Since TyphonQL queries are only executable on polystores precisely specified using TyphonML and TyphonDL, dedicated compilers/interpreters exploit this rich structural and semantical information to type-check and transform TyphonQL queries to high-performance native queries, and APIs that support advanced features such as prefetching and lazy loading to accommodate different usage scenarios.

The execution of TyphonQL queries will lead to the generation of events (also referred as "triggers" in the databases domain). These events are consumed by a high-performance framework for processing data access/update events to facilitate orthogonal real-time monitoring and predictive Analytics.

Finally, with information gathered from TyphonML, TyphonDL and the analytics components are used as input to the Evolution Engine which is responsible for evolving the organisation and distribution of data in hybrid polystores, as well as providing tools for monitoring the use of polystores to inform the evolution process.

III. BACKGROUND ON DEPLOYMENT TECHNOLOGIES

An application software (application for short) is a single or a group of software programs designed for end-users. In the context of deployment modelling in this document, the individual software programs that comprise an application will be referred to as services. Fig. 2 illustrates a simple example of an application named Location Management that consists of two services: A database service based on MariaDB and a Spring Boot service that communicates to the database service.

*A. Cloud Platforms*

Cloud platforms are platforms on the internet that provide cloud computing services and offer computation power, database storage, content delivery and other functionalities to businesses to support their product development. The cloud infrastructure is maintained by the platform provider and not by the individual platform user, which allows businesses and other application developers to focus completely on the product they are creating without any concerns on the underlying infrastructure to run their applications.

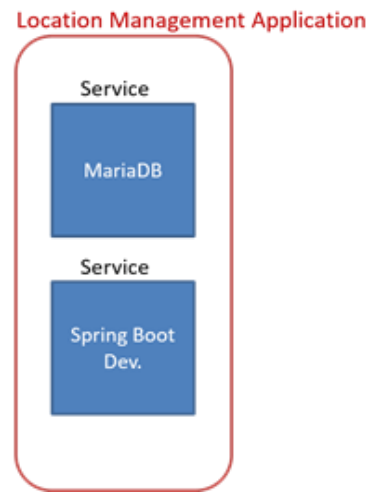

Fig. 2. An example application

Cloud platforms offer a wide range of benefits from providing flexibility in terms of 1) the scaling of infrastructure on demand in order to accommodate for varying workload; 2) public, private or hybrid storage options to meet the required security standards; 3) tool selection, to allow for accessibility of applications and data virtually from any device connected to the internet. This makes cloud platforms increasingly popular and useful.

Amazon Web Services [1], Microsoft Azure [2], Google Cloud[3], IBM Cloud[4] and Oracle Cloud[5] are amongst the most leading cloud platform providers.

*B. Containerised Applications using Docker*

A container is a standard unit of software that packages up code and all its dependencies, so the application runs quickly and reliably from one computing environment to another.

A Docker[6] container image is a lightweight, standalone, executable package of software that includes everything needed to run an application: code, runtime, system tools, system libraries and settings [5]. Contrary to virtual machines, multiple containers can share the OS kernel with other containers, thus taking up less space.

Images can be pulled from Docker-Hub [7], company specific docker registries either local or externally accessible (authentication with apache and nginx possible) or the Docker Trusted Registry (DTR), which is the enterprise-grade image storage solution from Docker.

Running multiple containers can be configured with the tool Docker-Compose [8]. All needed parts of an application (services) are defined in a YAML[9] file, which is used to create and start all needed containers with a single command.

In Docker-Compose, it is possible to specify the interplay between components in a deployment configuration – either linking to other services inside or even outside the `docker-compose.yaml`. Deployment properties (e.g., the computing power/number of CPUs to be used, amount of memory to be used, path to persistent storage, shared networks, etc.) can easily be added and edited.

---

[1] https://aws.amazon.com
[2] https://azure.microsoft.com
[3] https://cloud.google.com
[4] https://www.ibm.com/cloud/
[5] https://cloud.oracle.com
[6] https://www.docker.com/
[7] https://hub.docker.com/
[8] https://docs.docker.com/compose/
[9] https://yaml.org/

The deployment of the explanatory Location Management application on a cloud platform can be exemplified as in Fig. 3.

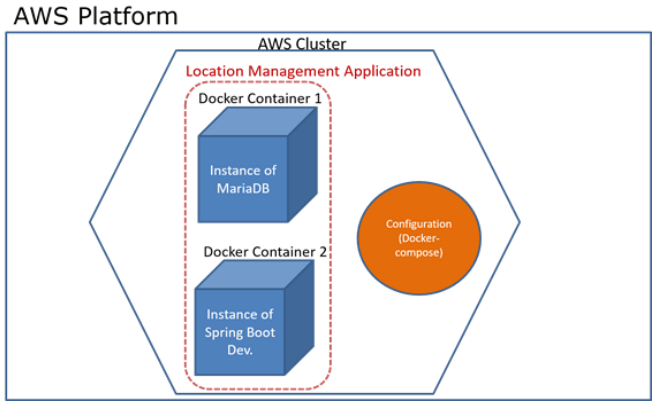

Fig. 3. Deployment example for an application

Each service in the Location Management application (the database service based on MariaDB and the Spring Boot service) in Fig. 1 is deployed as service instance separately in a Docker container on the AWS cloud platform.

Fig. 4 shows the corresponding deployment configuration specification for this application, where the Docker Container 1 is named as mariadb and the Docker Container 2 is named as locman. A service codifies the way an image runs. For example, the service mariadb uses the standard mariadb image from the Docker-Hub

https://hub.docker.com/_/mariadb/

plus a configuration-file, which changes the default database charset to utf8. Both services use gitlab as private container registry.

```
version: '3.5'
services:
  mariadb:
    image: gitlab.dummy.de:5555/atb/docker-base-images/mariadb:10.3.7-utf8
    container_name: mariadb
    environment:
      - MYSQL_ROOT_PASSWORD=geheim #[...]
    volumes:
      - /opt/locman-staging/db:/var/lib/mysql
    networks:
      - locman

  locman:
    image: gitlab.dummy.de:5555/europool/locman/location-mgmt-datastore:0.2.0-SNAPSHOT
    container_name: locman
    entrypoint: [
      '/maven/wait-for.sh', #[...]
    ]
    depends_on:
      - mariadb
    environment:
      - MARIADB_HOST=mariadb #[...]
    ports:
      - 8086:8086
    networks:
      - locman

networks:
  locman:
    name: locman
```

Fig. 4. docker-compose.yml example

### C. Container Management using Docker

In large-scale applications comprising hundreds of containers spread across multiple hosts, containers need to be managed and connected to the outside world for tasks such as scheduling, load balancing, and distribution. Docker images can be deployed and managed by container management solutions, such as Apache Mesos[10], Docker Swarm[11], or Kubernetes[12].

In TyphonDL, Kubernetes will be used as a container management system in the deployment of hybrid polystores [9], which can be extended to other container management systems in the future if desired. Kubernetes is an open source tool to orchestrate and manage containers. Furthermore, Kubernetes has good compatibility with Docker. It has been developed by Google and is one of the most used instruments for this purpose. Kubernetes allows removing many of the manual processes involved in the deployment and scalability of containerized applications and manage easily and efficiently clusters of hosts on which containers are executed.

### IV. PROPOSED CONCEPT FOR A HYBRID POLYSTORE DEPLOYMENT MODELLING LANGUAGE

While TyphonML models represent the high-level infrastructure of a hybrid polystore in terms of the conceptual entities to be managed and the corresponding database systems that are involved, TyphonDL models represent the deployment infrastructure of that polystore in terms of the specific cloud platform and deployment tools employed. The general approach of how TyphonDL is used for modelling the deployment infrastructure of a hybrid polystore is illustrated in Fig. 5. A TyphonDL model requires two sources of input:

- A TyphonML model, from which database specific information is translated for the TyphonDL model (e.g., which are the database systems that are used to manage the modelled data entities and relationships).
- Deployment specific values that instantiate configuration parameters, which generate a ready-to-use configuration file for the actual deployment task on a cloud platform.

The deployment specific values can be supplied by a modeller or a user of the polystore in textual form directly in the TyphonDL model itself or via a graphical editor that provides configuration specific input automatically in the TyphonDL model.

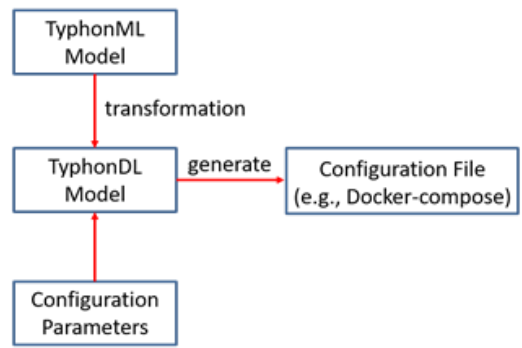

Fig. 5. TyphonDL approach

The remaining of this section is organized as follows: in the next Section the transformation from TyphonML models to TyphonDL models is elaborated in further detail. Followed by a Section about configuration specific input that is required for the creation of TyphonDL models is described. Finally, the metamodel of TyphonDL and its initial implementation

---

[10] http://mesos.apache.org/
[11] https://docs.docker.com/engine/swarm/
[12] https://kubernetes.io/

including tool support for TyphonDL are presented respectively.

*A. TyphonML to TyphonDL Transformation*

Recalling that a TyphonML model is a high-level specification of the database infrastructure of a hybrid polystore, TyphonDL follows the principle that each individual database installation will be deployed on a cloud platform in a separate container. Furthermore, based on what type of database systems are being modelled, the configuration parameters for those specific database systems will vary in the deployment model. In order to automatically generate the respective set of configuration parameters for different types of databases, the transformation of a TyphonML model will extract the database type for each database declaration in the TyphonML model and add this information in the TyphonDL model respectively.

*B. Configuration Parameters*

In order to generate a ready-to-use configuration file for deployment of a polystore on the cloud, the following parameters are necessary to be provided either by the modeller or the polystore user:

- Specific cloud platform provider: The modeller/user can specify a specific provider in the TyphonDL model such as AWS, Google Cloud, Microsoft Azure, etc.

- Specific container format: The modeller/user can specify which containerization technology should be used such as Docker, rkt[13], Virtual Box[14], VMWare[15].

- Specific of DBMSs: Based on the type of database retrieved from a TyphonML model, the modeller/user can choose a specific database system such as MariaDB[16] or MySQL[17] for a relational database, MongoDB[18], CouchDB[19] or RavenDB[20] as a document-based database, HBase[21], Hypertable[22] or Cassandra[23] as a column-based database, and Neo4j[24], AllegroGraph[25] or InfoGrid[26] as a graph-oriented database.

- Storage space for each cluster: Depending on the type of cloud platform provider, the modeller/user can specify in the TyphonDL model how much storage space is required for each cluster on the platform used for the deployment.

- Other platform dependent configuration options, such as computing power (number of CPU/GPUs) or amount of memory (RAM) can be specified by the modeller/user in the TyphonDL model.

- Database specific variables such as user name and password of an administrative user.

*C. TyphonDL Metamodel*

This section introduces the metamodel that formalizes the concepts that were analysed within the context of deployment technologies, which constitute the language primitives of TyphonDL. Meta-classes are shown in Fig. 6 and described below:

The object DeploymentModel represents the root container of each TyphonDL specification and consists of two distinct elements:

- *Metamodel*: It represents the collection of all relations between TyphonDL models.

- *Model*: It represents the collection of language-level concepts that are used to create a TyphonDL model.

Objects in TyphonDL models are categorized in three groups as follows:

- *Type*: It represents the collection of all types that are used in TyphonDL deployment models.

- *Deployment*: It represents the set of all concepts that are used in TyphonDL deployment models.

- *Software*: It represents the type of software implementations that are being deployed.

Types in a TyphonDL model are further sub-classified as follows:

- *PlatformType* represents the set of different types of platforms that can be used in a deployment task in the cloud.

- *ContainerType* represents the set of different types of containerization software that can be used in a deployment task.

- *DBType* represents the set of different database implementations such as MariaDB, MongoDB etc.

In the early design of TyphonDL one deployment model is associated with exactly one type of platform.

Deployments in a TyphonDL model are nested elements that begin with top-level declaration of a *platform*.

*Platform* is a named element and is typed by a platform type defined by PlatformType. It permits to model an individual platform space on a specific platform provider. It consists of a list of cluster declarations.

*Cluster* is a named element and represents a cluster that is running in a specific platform. It consists of a list of application declarations.

*Application* is a named element that represents a software-based application that is possibly composed of several smaller software components that are deployed in individual containers.

---

[13] https://coreos.com/rkt/
[14] https://www.virtualbox.org/
[15] https://www.vmware.com/
[16] https://mariadb.org/
[17] https://www.mysql.com/
[18] https://www.mongodb.com/
[19] http://couchdb.apache.org/
[20] https://ravendb.net/
[21] https://hbase.apache.org/
[22] http://www.hypertable.org/
[23] http://cassandra.apache.org/
[24] https://neo4j.com/
[25] https://allegrograph.com/
[26] https://infogrid.org/

*Container* is a named element that represents a virtual machine or a container based on technologies. It consists of a list of Properties that represent container specific parameters.

*Property* is a set of three different kinds of configuration declarations in the form of:

- key and value
- key and list of values
- key and array of values

pairs, and permit to represent properties specified in a configuration tool such as the image a container is running from, which network the container is connected to, the volume of a container, etc.

In TyphonDL, software implementations that are being deployed are classified into two types of implementations: database implementations and non-database implementations.

*Database* is a named element that specifies deployment specific properties of a concrete database, such as the image of the database system it belongs and authentication credentials. It is typed by the database system it belongs to.

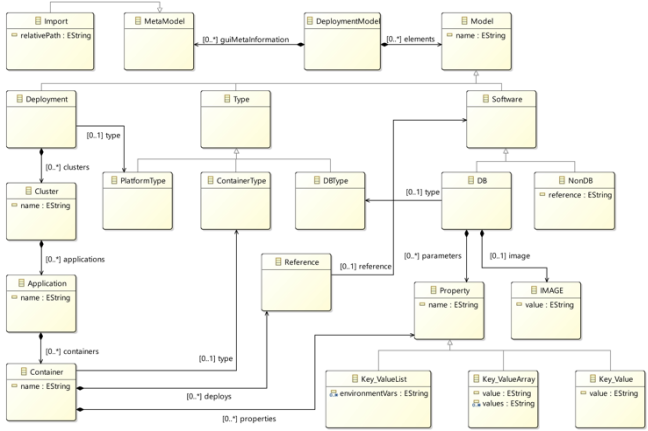

Fig. 6. TyphonDL concept hierarchy (class diagrams)

## V. PROTOTYPE IMPLEMENTATION

The implementation of TyphonDL is given as an EMF/Ecore [6] model. In parallel to this implementation, an initial version of a graphical editor is currently being implemented, which will allow polystore designers and users to create and/or to edit TyphonDL models (see following sub section B).

The initial version of the TyphonDL textual editor has been developed in XText[27], an Eclipse project for the design and development of domain-specific languages. The implementation of TyphonDL in XText follows from the grammar definition for the concrete syntax for TyphonDL. Upon the compilation of the TyphonDL grammar, XText generates an Ecore-model [7], the internal infrastructure for the parsing, linking, type-checking of TyphonDL models written textually in XText.

The eclipse project Sirius[28] is able to use the generated Ecore-model to create a Treeview of the TyphonDL model (see Fig. 6, left) and class diagrams of the grammar [8].

---

[27] https://www.eclipse.org/Xtext/

### A. TyphonDL Concrete Syntax

In this section the concrete syntax of TyphonDL is described using an example TyphonDL model as illustrated in Fig. 7.

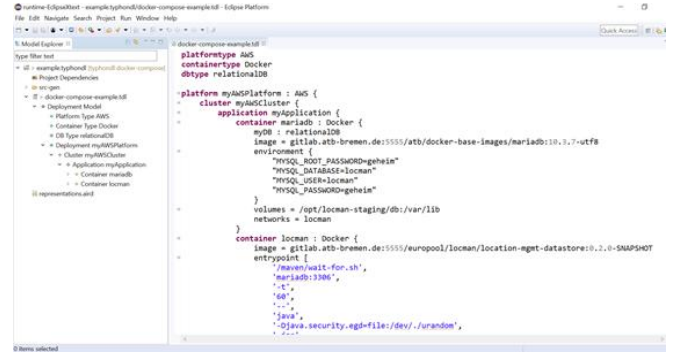

Fig. 7. TyphonDL example for docker-compose

The deployment example used in this TyphonDL model uses the Amazon Web Services (AWS) as the platform provider, Docker as the containerization technology and relational database as the database type. More specifically, this is modelled by the following type declarations:

```
platformtype AWS
containertype Docker
dbtype MariaDB
```

where `AWS` is declared as the platform type, `Docker` is declared as the container type and `MariaDB` is declared as the database type.

The declaration below is the specification of container-relevant properties of a MariaDB database named `locmandb`. It specifies the image that is used for the deployment of this database and authentication properties such as root password, user and user password.

```
database locmandb : MariaDB {
    image=gitlab.atb-bremen.de:5555/atb/docker-base-images/...
    MYSQL_ROOT_PASSWORD=geheim
    MYSQL_DATABASE=locman
    MYSQL_USER=locman
    MYSQL_PASSWORD=geheim
}
```

Next is the declaration of a specific AWS-platform, which is named `myAWSPlatform` and typed by `AWS`. `myAWSPlatform` consists of a cluster declaration named `myAWSCluster`, which consists of an application declaration named `myApplication`. Inside the application `myApplication` two Docker containers are modelled.

```
platform myAWSPlatform : AWS {
  cluster myAWSCluster {
    application myApplication {
      container myContainer : Docker {
        deploys locmandb
      }
      volumes = /opt/locman-staging/db:/var/lib
      networks = locman
    }
```

For the purposes of readability in the section, it is sufficient to explain the first container declaration, which is named `myContainer` and typed by the container type `Docker`. The container `myContainer` consists of reference to the database `locmandb` from which the image and database specific properties are obtained.

---

[28] https://www.eclipse.org/sirius/

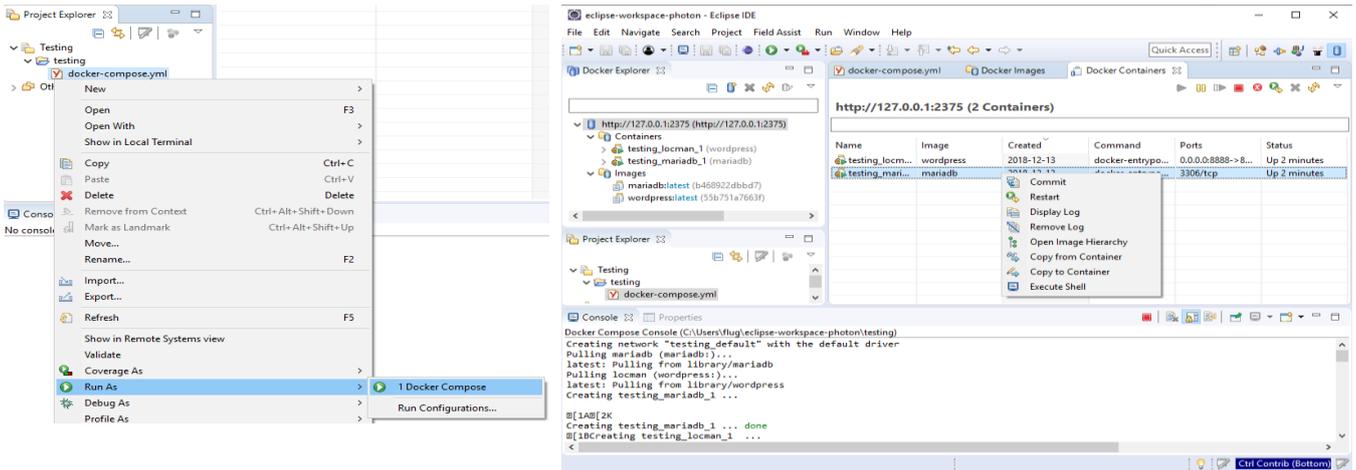

Fig. 9. Docker tooling example

*B. TyphonDL Creation and Editing*

For creating TyphonDL-models from TyphonML-models and editing them, a TyphonDL wizard and editor is being implemented as Eclipse plugins. The TyphonDL wizard retrieves a list of needed uniquely named databases and their types (i.e. relational, document etc.) from a given TyphonML model. In a GUI the specific DBMSs can be chosen from a list of supported systems for each required type. The technology for deployment (e.g. Docker Compose) has to be selected from a list as well. Other required user-input can be given through:

- model files that were created before, e.g. an image-specific .tdl file with all required information,

  -or-

- a GUI with DBMS- and technology-specific text fields. Fig. 8 shows the editing of a specific MongoDB container for Docker Compose. The editor can be used by regular users to change deployment specification without any knowledge of the TyphonDL grammar.

  -and/or-

- Modifying the generated DL model file directly. The editor shown in Fig. 8 is for expert users who are well acquainted with the TyphonDL grammar. Xtext provides features such as syntax highlighting and auto completion. The Sirius-plugin automatically generates the tree view in the Model Explorer shown on the left side of Fig. 8.

*C. Automated Generation of Configuration Scripts*

The generation of deployment scripts (e.g. the previously mentioned docker-compose.yml) out of a TyphonDL model is developed with Acceleo[29] an open-source project integrated in the Eclipse IDE. Through the specification of templates and referencing the Xtext model (provided in EMF format), files can be generated:

```
[template public generateDeployment(aDeploymentModel :
DeploymentModel)]
[for (aCluster : Cluster |
aDeploymentModel.getAllClusters())]
[for (aApplication : Application |
aCluster.getAllApplications())]

[file (aApplication.name + '/docker-compose.yml',
false, 'UTF-8')]
version: '3.7'
services:
[for (aContainer : Container |
aApplication.getAllContainers())]
    [aContainer.name/]:
```

...

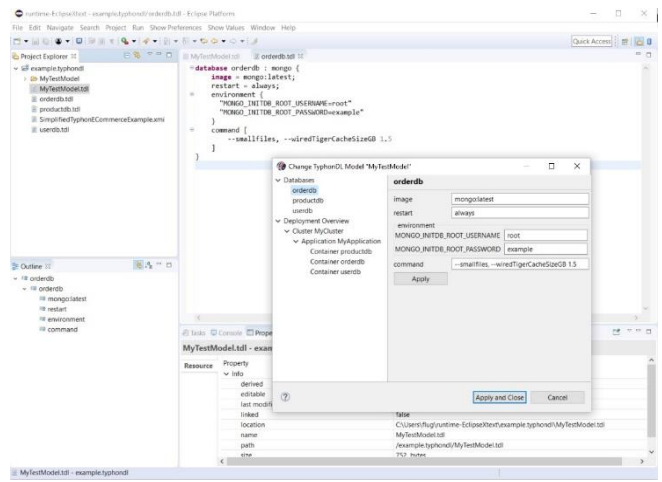

Fig. 8. GUI with text fields

Because of Acceleo's ability to produce any kind of source code or text from a given model, future extensions to the generator to support other technologies (different containerisation, Kubernetes etc.) are straightforward.

*D. Running Applications from Eclipse IDE*

Using Docker Tooling for Eclipse[30], containers can be created, started and stopped directly from the Eclipse-IDE (see Fig. 9).

VI. APPLICATION OF DEPLOYMENT LANGUAGE

One of the real-world applications of TyphonDL is a use case in automotive industry. The company driving this use case is one of the largest automobile manufacturers globally, and intends to radically reinvent its state-of-the-art services for commercial users. A key objective of the case study is to exploit available big data from smart vehicles and other sources such as geographic data, weather data, and data from a Content Management System (CMS), to create new business models and innovation scenarios for improved product quality, design and production cost-efficiency as well as customer satisfaction. The company already offers a comprehensive range of services around marketed products. In addition to the currently delivered services such as basic customer order management support, leasing services etc., the company intends to offer high quality data for the creation of new advanced products and customer services through the new proposed solution, by the ability to integrate and process data from diverse data stores.

---

[29] https://www.eclipse.org/acceleo/

[30] https://marketplace.eclipse.org/content/eclipse-docker-tooling

To further strengthen the customer relationship the company intends to foster the acceptance of innovative mobility concepts, by cloud based services supporting predictive vehicle maintenance by forecasting the behaviour of vehicle systems, enabling an efficient use of energy and transport capacities, and simultaneously supporting a data feedback loop to improve product quality, design and production cost-efficiency. Innovative smart vehicles, expected to be massively introduced in the next few years, offer opportunities to build new services and new business models. The addressed application scenario concerns progressive and intensive interaction concepts of smart vehicles with the outside world, enabling to offer a large amount of services by using the wide spectrum of information that can be provided by a vehicle. With such an approach, the potentials of the large amount of information inside the smart vehicle (using the vehicle as a Rolling Sensor), which may have huge utilization potentials for the outside world, no longer remain unused. Especially the integration of vehicle data with other data stores (geographic data, weather data, data from CMS, etc.) is of high interest. This will allow the company to support customers to accomplish an optimal use of vehicles (e.g. offer predictive vehicle maintenance services) with respect to customer-specific usage profiles and vehicle conditions (e.g. brake pad wear, battery condition etc.), as well as to use vehicle information to improve the design, production process and quality of product (e.g. detection of fault patterns for the identification of weak points in design). Research and Technological challenges in this use case include:

- Integration of datasets residing in different hybrid data stores with diverging structures (relational, file-based, graph and document databases).
- Introduction of advanced querying and evolving scalable architectures for persistence and real-time analytics and monitoring of large volumes of hybrid data (vehicle data, OEM data sources, 3rd party data sources).
- Data exploration and visualization services for integration of vehicle data with other data stores to gain further knowledge, enabling the development of new advanced products and customer services.
- Online pre-processing of data inside the vehicle.

In order to fulfil these expectations the volumes of data and frequency of data harvesting that can be generated by a fleet of cars need to be analysed. In this field, the main technical characteristics of a car are:

- One or multiple networks embedded (4 to 12 CAN buses, MOST, FlexRay, Ethernet)
- High-frequency of data broadcast (≈4000 signals, up to 100 values per second)
- Diagnosis information that can be queried from up to ≈50 ECUs (Electronic Control Units) (up to 1024 signals per ECU)
- Internal calculated values that can be queried from up to ≈50 ECUs (up to 22,000 high frequency signals per ECU)
- Decentralized network of information sources

Given these features, even a small fleet of 1,000 smart vehicles can produce a data volume of up to 400 TB/day, offering a wide spectrum of information with its thousands of signals from diverse sensors and ECUs per vehicle. Precisely, the following table describes the expected data volumes in this scenario:

Table 4: Volumes of data in case study

| Data Source | Volume per car | 1000 car fleet | 1year @1000 cars |
|---|---|---|---|
| 4-12 CAN buses | ca. 12 GB/day actual data | 12 TB/day | 4.4PB |
| Optional MOST bus | ca. 210 GB/day expected bandwidth | 210 TB/day | 77PB |
| Optional FlexRay network | ca. 80 GB/day expected bandwidth | 80 TB/day | 30PB |
| Optional Ethernet network | ca. 80 GB/day expected bandwidth | 80 TB/day | 30PB |

In this context, it can be concluded that capturing all this information has a very high cost (professional data loggers can cost from ≈4,000 to 15,000 euros). Equally transmitting this data to a centralized backend can be really expensive (500 MB/vehicle operation hour – only from the CAN bus), and finally the cost of the long-term storage of all this raw data does not provide a neat benefit compared to the potential business value. The main challenges to be overcome with the TYPHON technology will be:

- Extension of existing solutions by advanced big data and mining techniques: Selection of suitable aggregation and processing approaches to handle the massive and volatile amount of vehicle data.
- Reduction storage and transmission costs by decentralizing data processing and compressing data transfer.
- Introduction of advanced querying and evolving scalable architectures for persistence and real-time analytics and monitoring of large volumes of hybrid data.

- Enrichment of vehicle data with further 3rd party OEM data sources (e.g. model information, supply chain details, etc.) and other data sources.

- Development of scalable data exploration and visualization services for analytics and handling of storage over ever-growing vehicle data to gain further knowledge.

- Adoption of machine learning algorithms to provide predictive and prescriptive capabilities for the creation of new services.

## VII. CONCLUSIONS

In this paper a concept for designing, deploying, querying, evolving and analysing hybrid persistence architectures is proposed, that fulfils growing scalability and heterogeneity requirements of organisations. The focus is put on new domain-specific language, named TyphonDL giving some insights on its underlying principles and the current status of implementation.

In the future, the development of constructs and the tools of all the languages need to be finalised and refined based on feedback while working on real case scenarios. Among others it is planned to include other container technologies and the execution of Data Definition Language commands, support for the definition of individual nodes and of standard configuration concepts (e.g. DB master-slave nodes, Elastic instances, etc.) using TyphonDL.

## ACKNOWLEDGMENT

This work is partly supported by the TYPHON (Polyglot and hybrid persistence architectures for Big Data analytics) project of European Union's Horizon 2020 Framework Program, under the grant agreement no. 780251. This document does not represent the opinion of the European Community, and the Community is not responsible for any use that might be made of its content.